\documentclass[conference,a4paper]{IEEEtran}
\usepackage[pdftex]{graphicx,xcolor}
\usepackage[fleqn]{amsmath}
\usepackage{newtxtext}
\usepackage[varg]{newtxmath}
\usepackage{subfig}
\usepackage[normalem]{ulem}
\usepackage{multirow}
\usepackage{dblfloatfix}
\usepackage{xcolor,cancel}
\usepackage{tikz}
\usepackage{algorithm}
\usepackage{algpseudocode}

\usepackage{booktabs} 

\useunder{\uline}{\ul}{}

\newcommand{\AmSLaTeX}{%
	$\mathcal A$\lower.4ex\hbox{$\!\mathcal M\!$}$\mathcal S$-\LaTeX}
\def\BibTeX{{\rmfamily B\kern-.05em
		\textsc{i\kern-.025em b}\kern-.08em
		T\kern-.1667em\lower.7ex\hbox{E}\kern-.125emX}}
\makeatletter
\def\tmpcite#1{\@ifundefined{b@#1}{\textbf{?}}{\csname b@#1\endcsname}}%
\makeatother

\title{Subjective Quality Assessment of Dynamic 3D Meshes in Virtual Reality Environment}
\author{
	\IEEEauthorblockN{Duc V. Nguyen$^1$, Nguyen Thi Quynh Ly$^2$, Truong Thu Huong$^2$}
	\IEEEauthorblockA{
		$^1$\textit{Faculty of Engineering, Tohoku Institute of Technology, Japan}\\
		$^2$\textit{School of Electrical and Electronic Engineering, Hanoi University of Science and Technology, Vietnam}}
}

\begin{document}
	\maketitle
	
	\begin{abstract}
		A dynamic 3D mesh is a key component in Virtual Reality applications. However, this type of content demands a significant processing resource for real-time rendering. To reduce processing requirements while preserving the user experience, adjusting the level of detail of 3D meshes based on viewing distance has been proposed. In this paper, we conduct an extensive subjective quality evaluation to investigate the effects of the level of detail and viewing distance on user perception of dynamic 3D meshes in a VR environment. Our evaluation results in a subjective dataset containing user ratings of 320 test stimuli generated from eight dynamic 3D meshes. Result analysis shows that it is possible to remove half of a mesh's faces without causing noticeable degradation in user Quality of Experience (QoE). An evaluation of popular objective quality metrics reveals that both 2D-based and 3D-based metrics have low correlation with subjective scores. Based on the subjective dataset, we develop a novel QoE prediction model that can accurately predict the MOS of a dynamic 3D mesh at a given level of detail and viewing distance. In addition, a QoE-aware resource allocation framework is proposed and evaluated under different resource constraints, showing significant improvement in the total QoE compared to conventional methods. 
	\end{abstract}
	\begin{IEEEkeywords}
		Virtual Reality, Dynamic 3D Mesh, Level of Detail, Viewing Distance, Quality of Experience (QoE)
	\end{IEEEkeywords}

\section{Introduction}
Thanks to the introduction of consumer-oriented Head-Mounted Displays (HMDs), advancements in graphics processing, and high-speed access networks, Virtual Reality (VR) has become widely accessible nowadays. A recent survey shows that the VR market is projected to reach 435 billion USD by 2030~\cite{VRMarketReport}. With the ability to provide the \textbf{immersive experience} to users, Virtual Reality can offer great benefits to many fields, including education~\cite{DucGCCE2024}, training~\cite{VRforTraining}, and healthcare~\cite{VRHealthcare}.
		
In VR applications, a realistic representation of 3D objects is crucial to provide an immersive experience to the users. 3D meshes are one of the most popular formats to represent 3D objects in a Virtual Reality environment. A 3D mesh represents an object by a set of vertices, faces, and their attributes, including textured maps and normal vectors. For a realistic representation of 3D objects, complex 3D meshes containing a large number of faces are often required. As VR applications usually contain many 3D objects, huge computational resource is needed to render the 3D meshes in real-time. Meanwhile, consumer-oriented VR devices are resource-constrained with limited processing power. For instance, standard standalone VR headsets such as Meta Quest are not equipped with a Graphical Processing Unit (GPU). Thus, a key challenge in Virtual Reality is how to optimize the processing of 3D meshes representing virtual objects on resource-constrained user devices. To deal with this challenge, previous works have proposed adjusting the level of detail (LoD) of a 3D mesh based on the distance between the object and the viewer~\cite{quang2023visibility,Quang2025}. The rationale is that the further away the 3D object is, the more difficult for the user to see the object’s details. Thus, it is possible to lower the object's level of detail to reduce the processing requirement without affecting the user’s experience.

For optimal generation and selection of the LoD of a 3D mesh, it is important to understand how the user perceives the quality of a dynamic 3D mesh at different levels of detail and viewing distances in a VR environment. Despite a large amount of work on quality assessment of graphical 3D content, there have been limited efforts regarding this problem in the literature. The impacts of the viewing distance and LoD on user perception of 3D meshes in a VR environment are studied in~\cite{DucIEICETrans2024,DucICIP2024}. However, these works consider static 3D meshes that do not contain the object's movement. Meanwhile, the majority of 3D meshes are dynamic with associated animations to represent the object's various movements. The work~\cite{Cao2020} considers the joint effect of viewing distance and compression on user perception of dynamic 3D meshes representing humans. However, the evaluation in~\cite{Cao2020} was conducted on a desktop environment.

\begin{table*}[t]
\caption{Comparison of this work and related works.}
\label{tab:related_work_comparison}
\resizebox{\textwidth}{!}{%
\begin{tabular}{|l|l|l|l|l|l|l|}
\hline
\textbf{Work} & \textbf{Source content} & \textbf{\#Stimulus} & \textbf{Content type} & \textbf{Considered   factors} & \textbf{Test   Evinronment} & \textbf{Format} \\ \hline
Cao et al., 2020~\cite{Cao2020} & 4 humans & 120 & Dynamic & Viewing Distance,   Compression & 2D Monitor & Point Cloud, Mesh \\ \hline
Minh Nguyen   et al., 2023~\cite{minhnguyen2023} & 2 humans & 18 & Dynamic & Viewing Distance,   Compression & Mixed Reality (Hololens 2) & Point Cloud \\ \hline
Duc Nguyen et al., 2024~\cite{DucICIP2024} & 5 objects & 80 & Static & Viewing Distance,   Level of Detail Simplication & Virtual Reality (Meta Quest 2) & Mesh \\ \hline
\textbf{Our's} & \textbf{8 objects} & \textbf{320} & \textbf{Dynamic} & \textbf{Viewing Distance,   Level of Detail Simplication} & \textbf{Virtual Reality} (\textbf{Meta Quest 2}) & \textbf{Mesh} \\ \hline
\end{tabular}%
}
\end{table*}

To fill in the gap in the literature, we conduct an extensive subjective quality evaluation of dynamic 3D meshes to investigate the effects of two key factors: 1) level of detail, and 2) viewing distance on the user's Quality of Experience (QoE) in a VR environment. The evaluation consists of 320 test stimuli generated from eight original dynamic 3D meshes with eight levels of detail and five viewing distances. Based on the subjective dataset, we develop a novel QoE model to predict the Mean Opinion Score (MOS) of dynamic 3D meshes with high accuracy. In addition, we introduce a QoE-aware resource allocation framework that utilizes the proposed QoE model to optimize user QoE under a resource-constrained environment. The proposed framework is shown to improve the overall user QoE by up to 30.5\% compared to conventional methods.

The remainder of the paper is organized as follows. The related work is described in Section~\ref{sec:relatedwork}. The subjective quality assessment is described in Section~\ref{sec:experiment}. Result analysis is performed in Section~\ref{sec:result_analysis}. The QoE prediction model is presented in Section~\ref{sec:QoE_model}. Section~\ref{sec:QoE_aware_resource_allocation} describes the QoE-aware resource allocation framework. Finally, the paper is concluded in Section~\ref{sec:conclusion}.

		\begin{figure*}[t]
			\centering
			\subfloat[M1]{
				\includegraphics[width=0.123\textwidth]{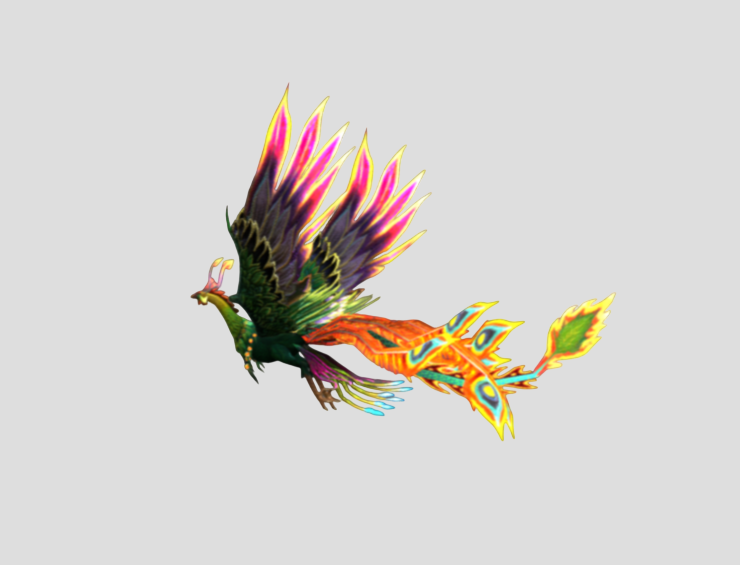}
			}
			\subfloat[M2]{
				\includegraphics[width=0.123\textwidth]{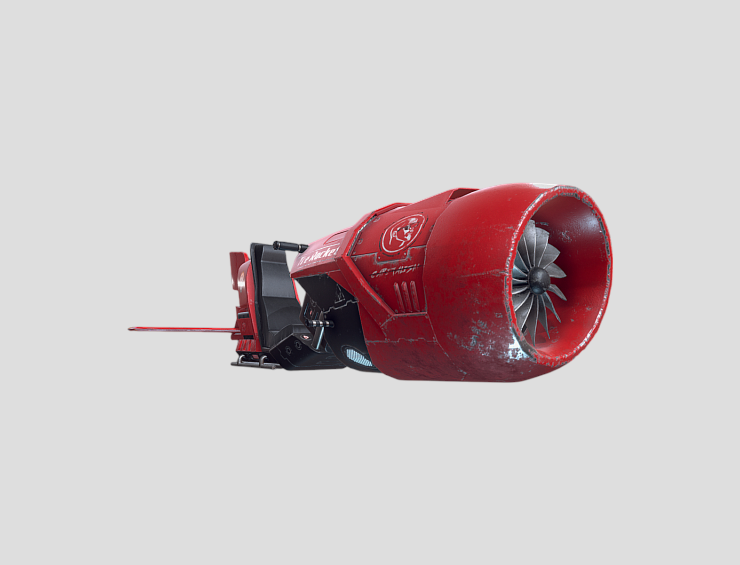}
			}
			\subfloat[M3]{
				\includegraphics[width=0.123\textwidth]{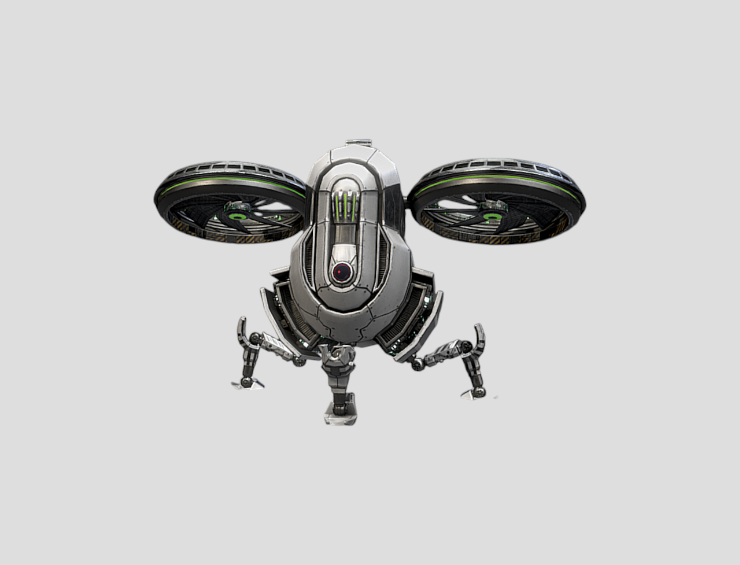}
			}
			\subfloat[M4]{
				\includegraphics[width=0.123\textwidth]{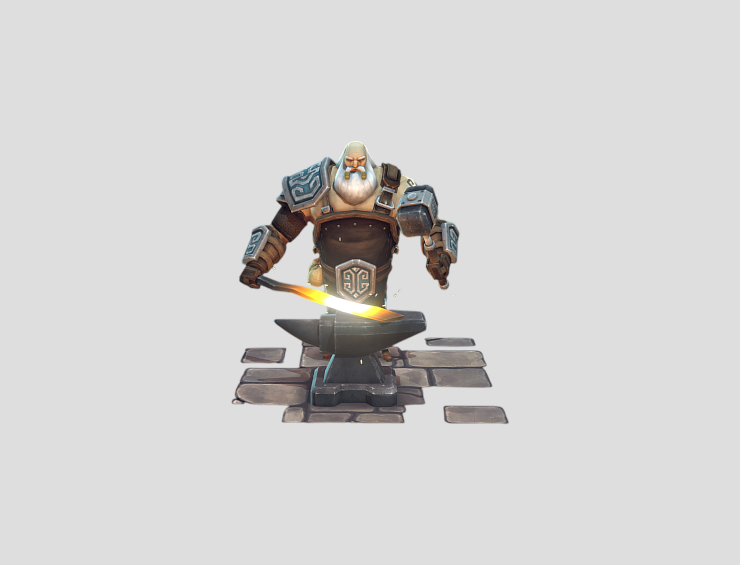}
			} 
			\subfloat[M5]{
				\includegraphics[width=0.123\textwidth]{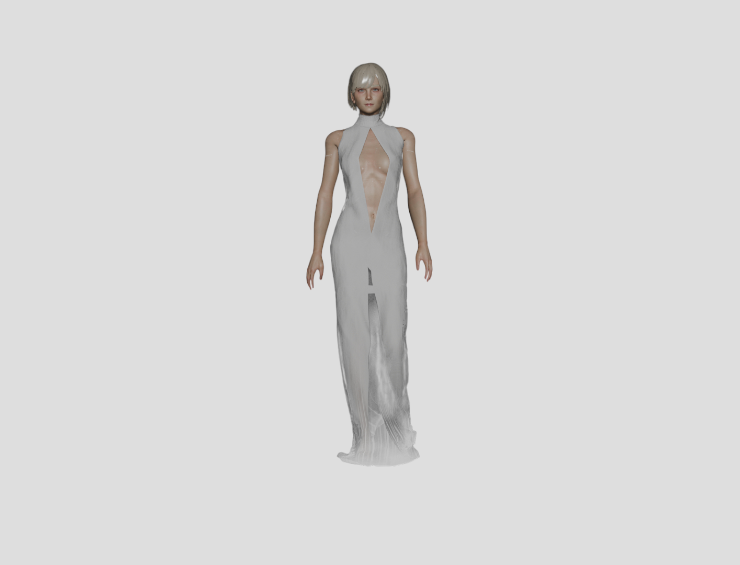}
			} 
            \subfloat[M6]{
                \includegraphics[width=0.123\textwidth]{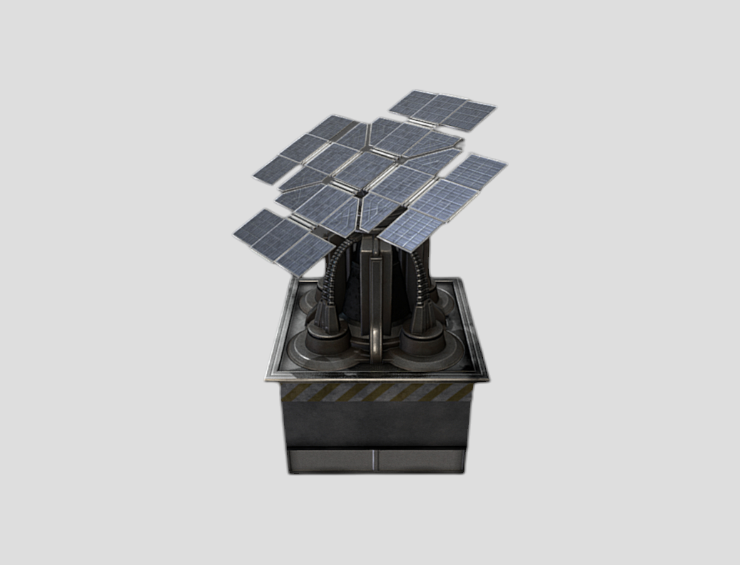}
            }
            \subfloat[M7]{
                \includegraphics[width=0.123\textwidth]{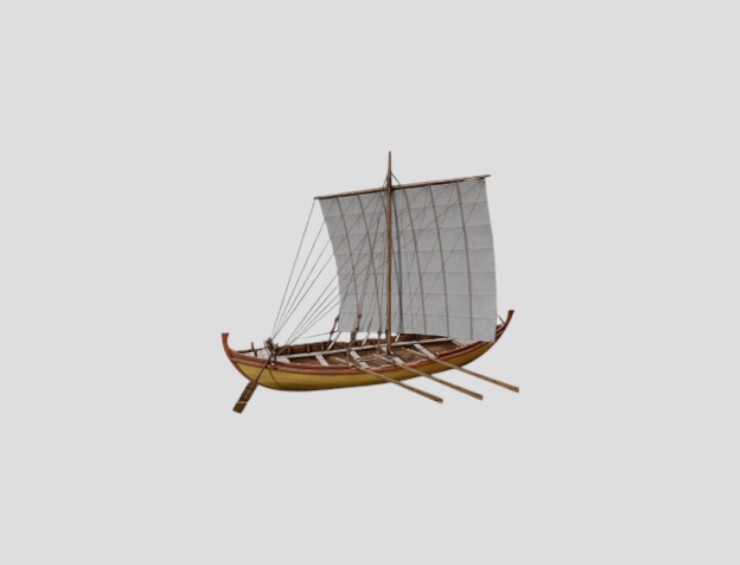}
            }
            \subfloat[M8]{
                \includegraphics[width=0.123\textwidth]{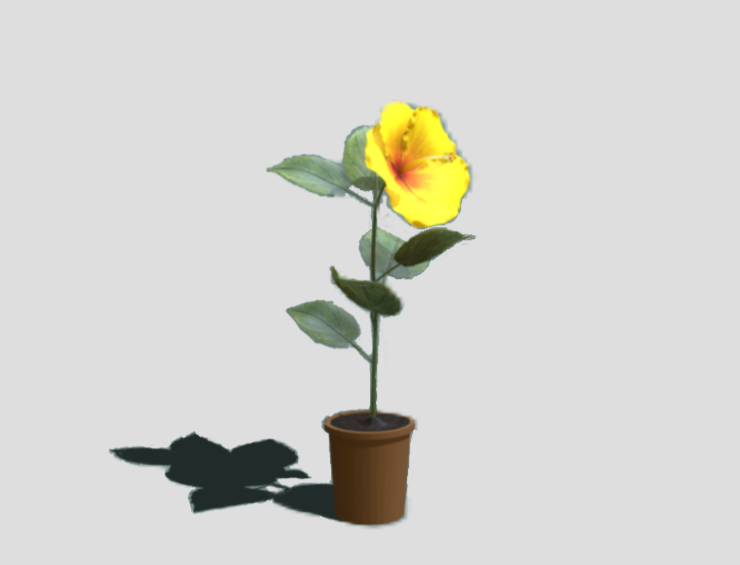}
            }
			\caption{Snapshots of eight dynamic 3D meshes used in the subjective experiment. Each mesh is associated with animations representing object movements.}
			\label{fig:3D_model_snapshot}
		\end{figure*}
		
\section{Related Work}\label{sec:relatedwork}
\subsection{3D Object Representation Methods}
There are several methods to represent a 3D object in VR applications. The conventional methods include 3D meshes~\cite{PolygonMesh} and point clouds~\cite{PointCloud}. A 3D mesh represents the shape of a 3D object using a set of polygons (usually triangles). Each polygon is defined by the vertices' 3D coordinates and a normal vector. Attributes include textured images that are mapped to the polygons to form the surface's appearance of a 3D object. In point cloud format, a 3D object is represented by a set of points in 3D space. Each point is defined by its 3D coordinates, colors, and normal data~\cite{PointCloud}. Recently, AI-based methods for 3D scene representation, such as Nerf~\cite{Nerf2020} and Gaussian Splashing~\cite{3DGS2023}, are gaining popularity. Unlike mesh and point cloud, AI-based methods learn a function that maps 3D coordinates and viewing direction to color and density. This allows the synthesis of an arbitrary view of a 3D object from a given viewing direction. Although offering several benefits, both point cloud and AI-based representations require significant computational resources for rendering. In this paper, we consider the 3D mesh format and reserve other formats for our future work.
	
\subsection{Subjective Quality Assessment of 3D Meshes}
In the literature, subjective quality assessment of 3D mesh has been conducted to study the impacts of various factors on user perception. Different content types (colorless~\cite{pan2005quality}, colored~\cite{nehme2023textured}),  distortion types (compression~\cite{Cao2020}, simplification~\cite{DucICIP2024}, sampling~\cite{Guo2016}), subjective test methodologies, content presentation methods, and test environments (desktop~\cite{Guo2016}, VR~\cite{DucICIP2024}, MR~\cite{minhnguyen2023}) have been considered.

Regarding distortion types, most previous work focuses on quantifying the impacts of distortions caused by compression, downsampling, geometry simplification, and noise on both geometry and texture data. In~\cite{pan2005quality}, it is shown that viewers are more sensitive to the distortion on the texture than to the distortion of the geometry. In~\cite{Guo2016,nehme2023textured}, the authors study the impacts of compression, simplification, and smoothing on geometry data, JPEG compression, and subsampling on texture data. It is shown that 3D meshes with complex textures are very sensitive to simplification, whereas highly curved models are sensitive to smoothing. It is also shown that the perception of geometry quantization artifacts is strongly dependent on the level of detail~\cite{nehme2023textured}. The effects of quantization and simplification on geometry and color data are studied in~\cite{Yana2021}. Environment-related factors such as lighting conditions~\cite{gutierrez2020quality} and light-material interaction~\cite{Vanhoey2017} have also been investigated.  

The test methodology is mainly based on existing image/video quality assessment methods, including single stimulus methods, in which participants see one test stimulus at a time and give a quality score~\cite{gutierrez2020quality,DucIEICETrans2024,Subramanyam2020,viola2022impact}, double stimulus methods in which participants view the original mesh followed by a distorted mesh, then rate the impairment level of the distorted mesh~\cite{DucICIP2024,nehme2023textured,Yana2021}. Two main approaches are used to present 3D mesh content to participants. The first approach is to render 2D images or videos of a 3D mesh based on predefined viewpoints and camera trajectories. The images/videos are then displayed on a 2D monitor~\cite{Guo2016,Cao2020,pan2005quality,Zerman2020,nehme2023textured}. In the second approach, 3D meshes are displayed directly in a VR/MR environment~\cite{gutierrez2020quality,minhnguyen2023,DucICIP2024,DucISM2024}. This approach better mimics the practical setting in VR applications and will be used in this study.

Despite a large amount of previous work, little efforts have been made to jointly quantify the impacts of level of detail and viewing distance on the user perception of 3D meshes in a VR environment. In~\cite{Cao2020}, the effects of compression and viewing distance on user-perceived quality of four dynamic 3D meshes are investigated. However, the evaluation in~\cite{Cao2020} is conducted on a desktop environment. In~\cite{minhnguyen2023}, the effects of viewing distance on user perception of two dynamic point clouds are studied in a Mixed Reality setting. However, the joint effect of viewing distance and quality degradation is not examined. In~\cite{DucICIP2024}, user perception of five static 3D meshes under different levels of detail and viewing distances in a VR environment is studied. However, unlike our work that considers dynamic 3D meshes, this work considers only static 3D meshes. A summary of the key differences between this work and related works is given in Table~\ref{tab:related_work_comparison}.

        \begin{figure}[t]
            \centering
            \includegraphics[width=\columnwidth]{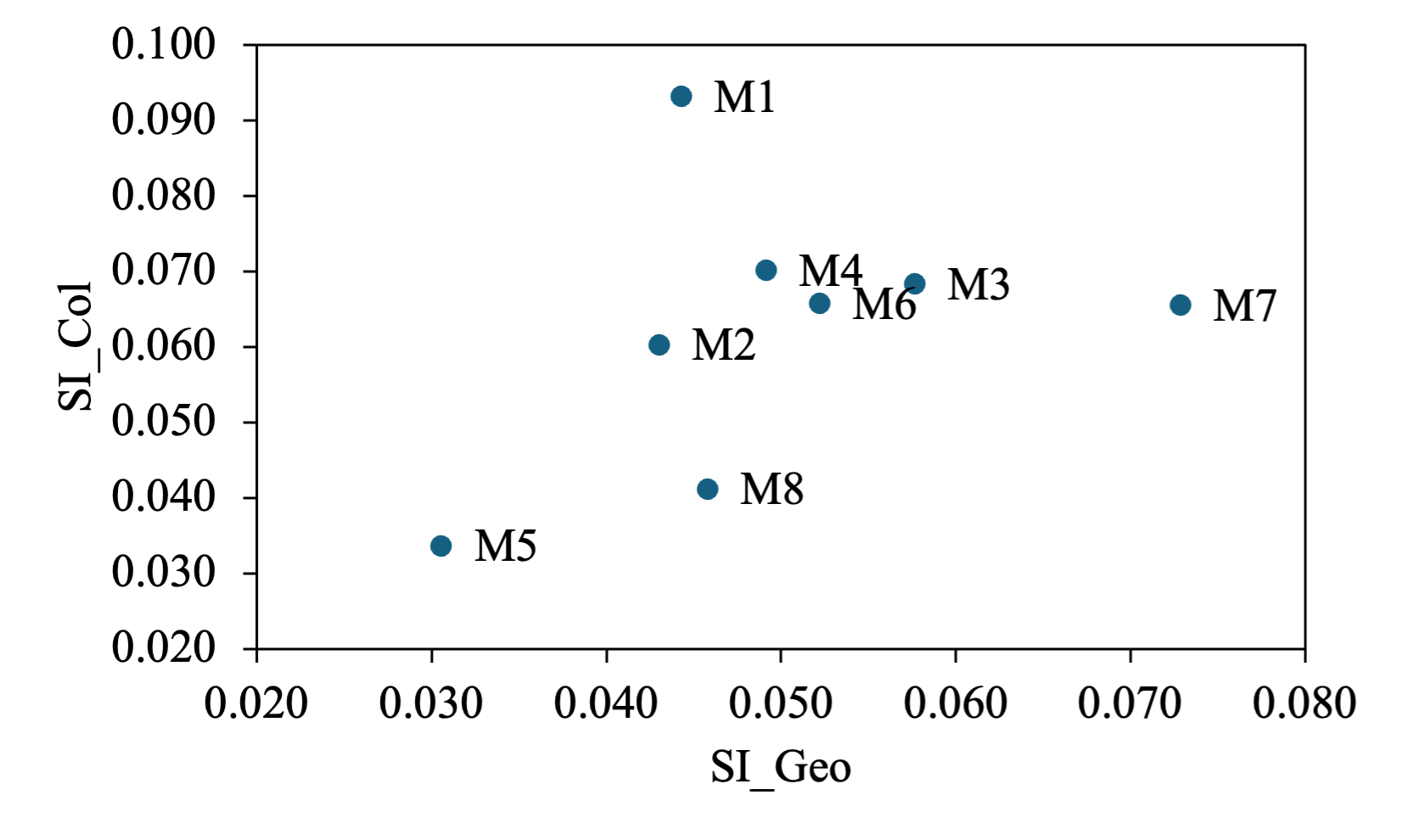}
            \caption{Geometry (SI\_Geo) and color (SI\_Col) spatial information of the considered dynamic 3D meshes.}
            \label{fig:mesh_SI}
        \end{figure}
        
\section{SUBJECTIVE QUALITY EXPERIMENT}\label{sec:experiment}
        \subsection{Test Stimuli Generation}
        
        In this study, we use eight dynamic 3D meshes from Sketchfab\footnote{https://sketchfab.com/}. All meshes are licensed under Creative Commons licenses. The snapshots of the 3D meshes are shown in Fig.~\ref{fig:3D_model_snapshot}. All of the meshes are artificial models created using modeling software with structured texture content and smooth motions. Each 3D mesh is associated with a 10-second-long animation. The complexity of the 3D meshes specified in geometry and color domain spatial information~\cite{nehme2023textured} is shown in Fig.~\ref{fig:mesh_SI}. It can be seen that the chosen meshes cover a wide range of both geometry and color complexity. 
        \begin{table}[]
			\caption{Number of faces of eight LoDs of the eight considered dynamic 3D meshes.}
            \centering
			\label{tab:LoD_version_vertice}
			\resizebox{\columnwidth}{!}{%
				\begin{tabular}{|l|l|l|l|l|l|l|l|l|}
					\hline
					\textbf{LoD} & \textbf{M1} & \textbf{M2} & \textbf{M3} & \textbf{M4} & \textbf{M5} & \textbf{M6} & \textbf{M7} & \textbf{M8}\\ \hline
					\textbf{Original} &4063	&25246	&32720	&6832	&77980 &16556  &66776  &221874 \\ \hline
                    \textbf{LoD1} &3250    &20193  &26149  &5457   &65122  &13237  &54106  &169489 \\ \hline
                    \textbf{LoD2} &2436    &15144  &19606  &4087   &52277  &9928   &41547  &127116 \\ \hline
					\textbf{LoD3} &2030	&12623	&16361	&3589	&46493  &8277   &35144  &105935 \\ \hline
                    \textbf{LoD4} &1625    &10094  &13063  &2719   &39429  &6620   &28805  &84744 \\ \hline
					\textbf{LoD5} &1218	&7570	&9826	&2104	&33333  &4963   &22482  &63557 \\ \hline
                    \textbf{LoD6} &812     &5045   &6526   &1349   &26588  &3307   &16159  &42366 \\ \hline 
					\textbf{LoD7} &406	&2522	&3326	&790	&20643    &1646   &9832   &21178 \\ \hline
					\textbf{LoD8} &202	&1260	&1703	&398	&11360    &825     &6775   &10588 \\ \hline
				\end{tabular}%
			}
		\end{table}
        
        To adjust the level of detail of a dynamic mesh, we use the \textit{Decimate} modifier of the Blender software that reduces the face count of a 3D mesh with minimal shape change by edge collapse~\cite{BlenderModifier}. In particular, eight LoDs are generated for each mesh corresponding to the percentage of removed faces set to 20\%~(LoD1), 40\%~(LoD2), 50\%~(LoD3), 60\%~(LoD4), 70\% (LoD5), 80\% (LoD6), 90\% (LoD7), and 95\% (LoD8). The number of faces of the LoDs of the considered 3D meshes is shown in Table~\ref{tab:LoD_version_vertice}. For the viewing distance, we consider five distance values of $\{4,8,12,16,20\}$ meters. The distances are selected so that the whole model is in the user's viewport at the shortest distance (4m). Also, the model's size in the viewport becomes relatively small at the furthest distance (20m). In total, our subjective experiment consists of 320 test stimuli (8 meshes $\times$ 8 LoDs $\times$ 5 distances).

        \subsection{Test Environment, Methodology, Procedure}
        
        To conduct the quality assessment in a Virtual Reality environment, we develop a Web-based VR test environment using the A-Frame framework\footnote{https://aframe.io/}. The test stimuli are stored on an HTTP-based Web server. The user agent (i.e., Web browser) preloads the test stimuli in advance to ensure smooth display of the 3D meshes. The test stimuli are displayed automatically in a random order. A rating interface is displayed after each test stimulus. For the viewing device, we use a Meta Quest 2 Virtual Reality headset with a per-eye resolution of 1832$\times$1920 and a refresh rate of 90 Hz.

        For the test methodology, we employ the double stimulus impairment scale (DSIS) method~\cite{itu_t_p_910_2023}. For a test stimulus, the corresponding original 3D mesh is displayed first, followed by the test stimulus. Afterward, the participant is asked to rate the quality of the test stimulus in comparison to that of the original 3D mesh on a 5-grade impairment scale: 5 (\textbf{Imperceptible}), 4 (\textbf{Perceptible, but not annoying}), 3 (\textbf{Slightly annoying}), 2 (\textbf{Annoying}), and 1 (\textbf{Very annoying}).

		\textcolor{black}{At the beginning of a test session, the participants are informed about the objective of the experiment. Then, they are trained to get used to the VR headset and controllers. The test stimuli are displayed in a random order to mitigate the ordering effect. The participants use controllers to give the score of each test stimulus, which is then automatically recorded by the evaluation system. Each test stimulus is evaluated by twenty participants. The Mean Opinion Score (MOS) of a stimulus is calculated as the average score of all participants.} 

		\begin{figure}[t]
            \centering
            \includegraphics[width=0.8\columnwidth]{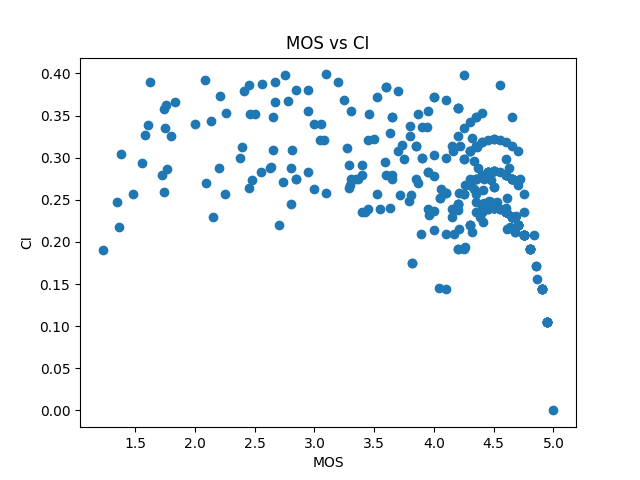}
            \caption{MOS values and 95\% Confidence Interval of test stimuli}
            \label{fig:ci}
        \end{figure}

		\begin{figure*}[t]
			\centering
			\subfloat[d=4m]{
				\includegraphics[width=0.3\textwidth]{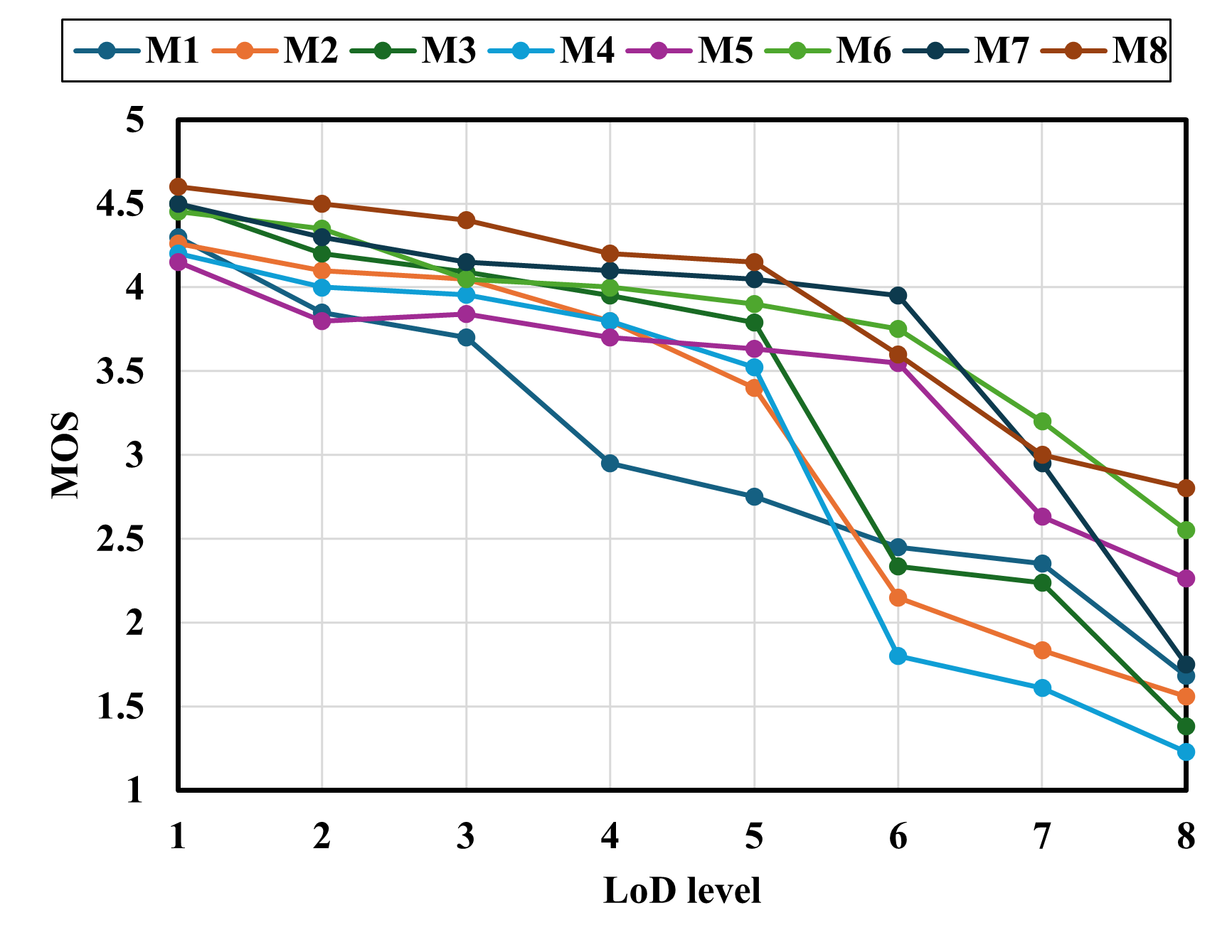}
			}
			\subfloat[d=8m]{
				\includegraphics[width=0.3\textwidth]{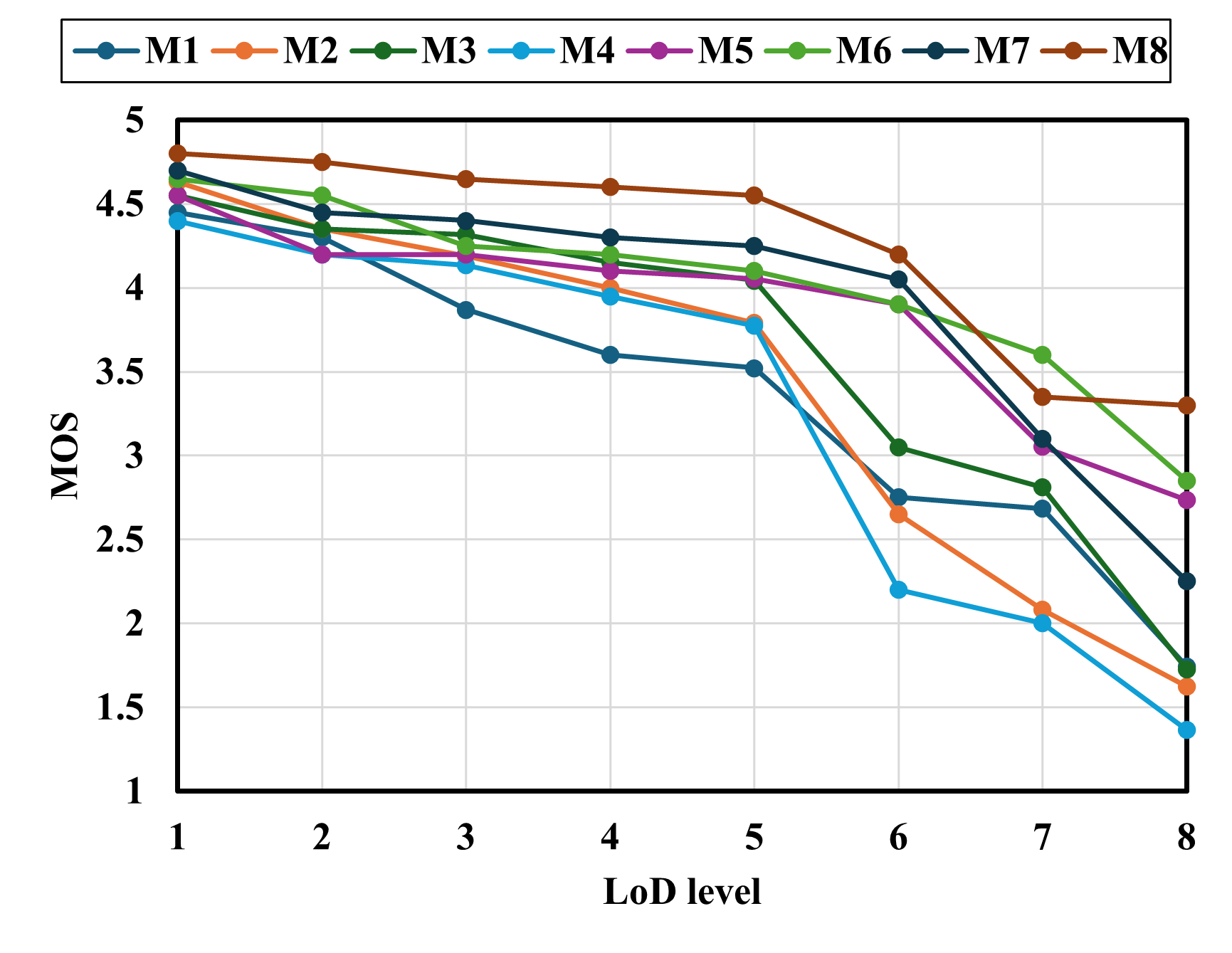}
			}
			\subfloat[d=12m]{
				\includegraphics[width=0.3\textwidth]{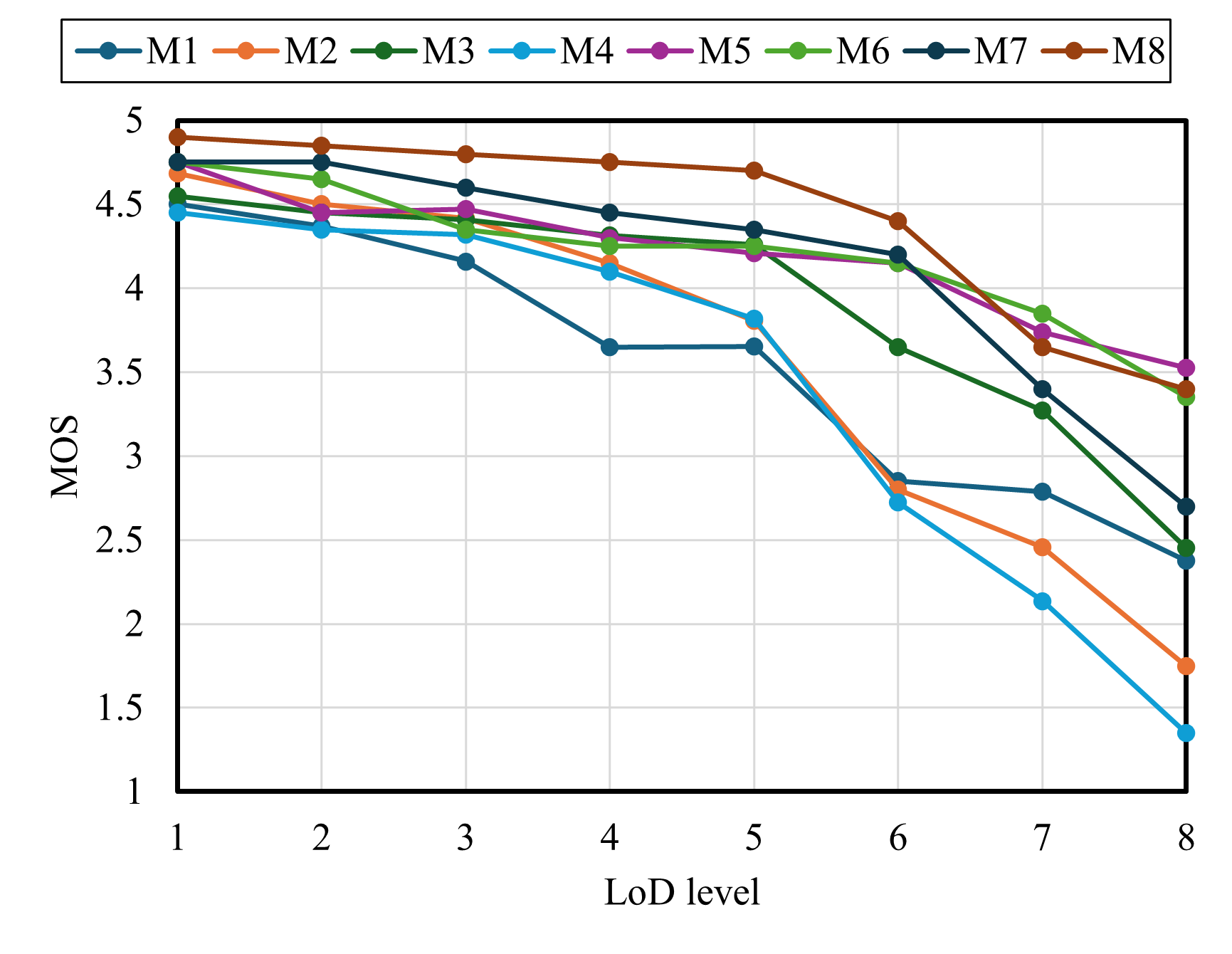}
			}\hfill
			\subfloat[d=16m]{
				\includegraphics[width=0.3\textwidth]{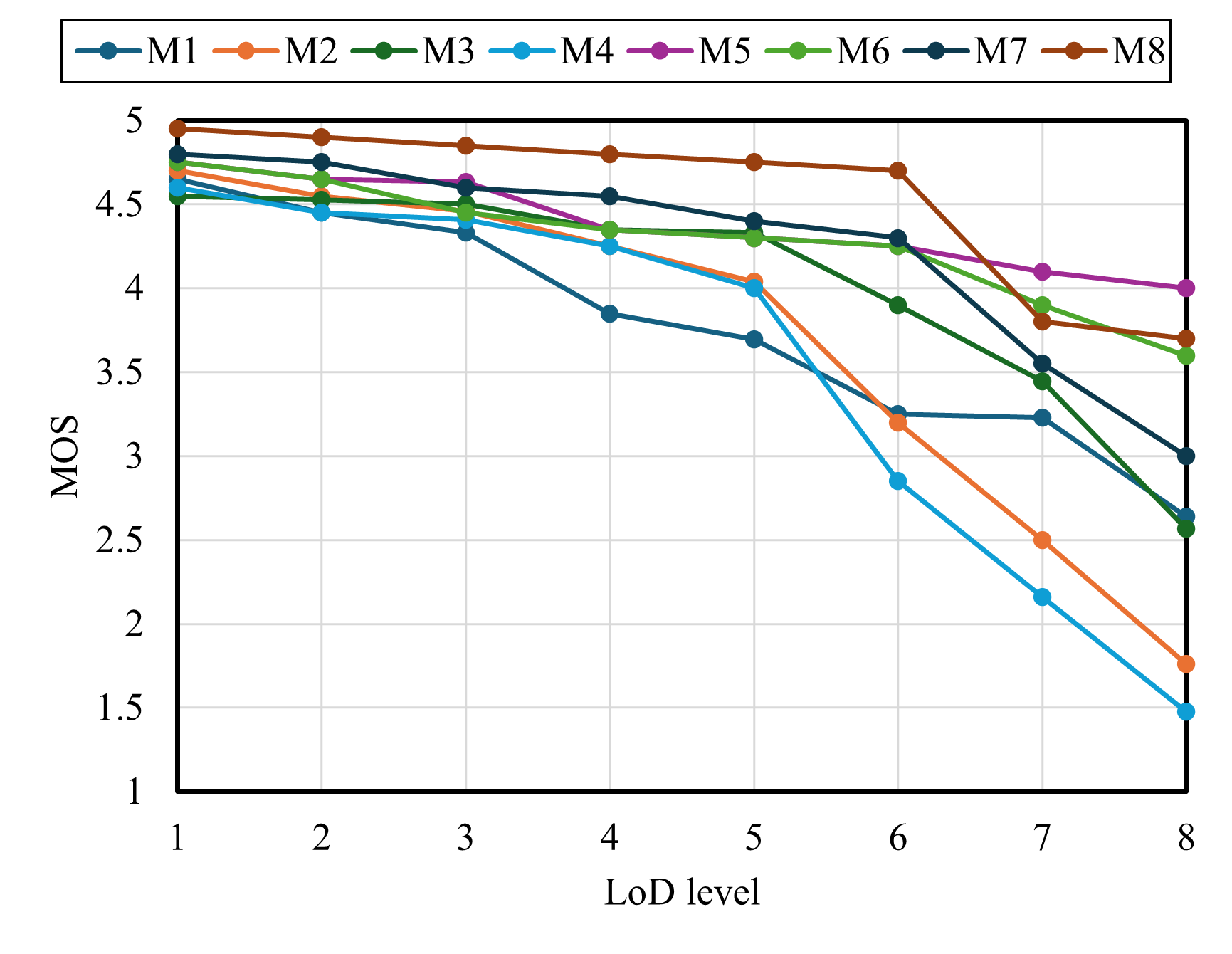}
			}
			\subfloat[d=20m]{
				\includegraphics[width=0.3\textwidth]{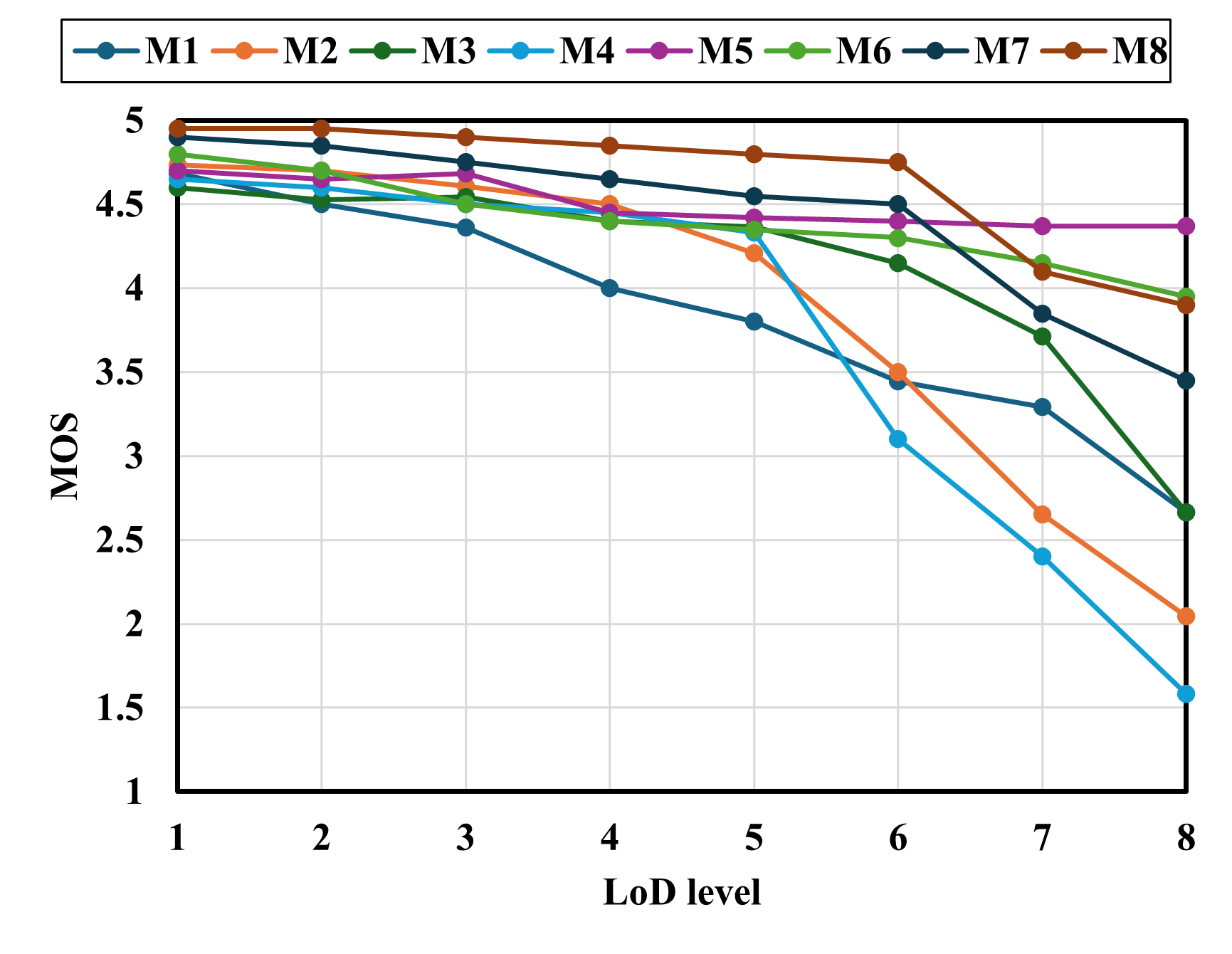}
			}
            \subfloat[Average over all viewing distances]{
				\includegraphics[width=0.3\textwidth]{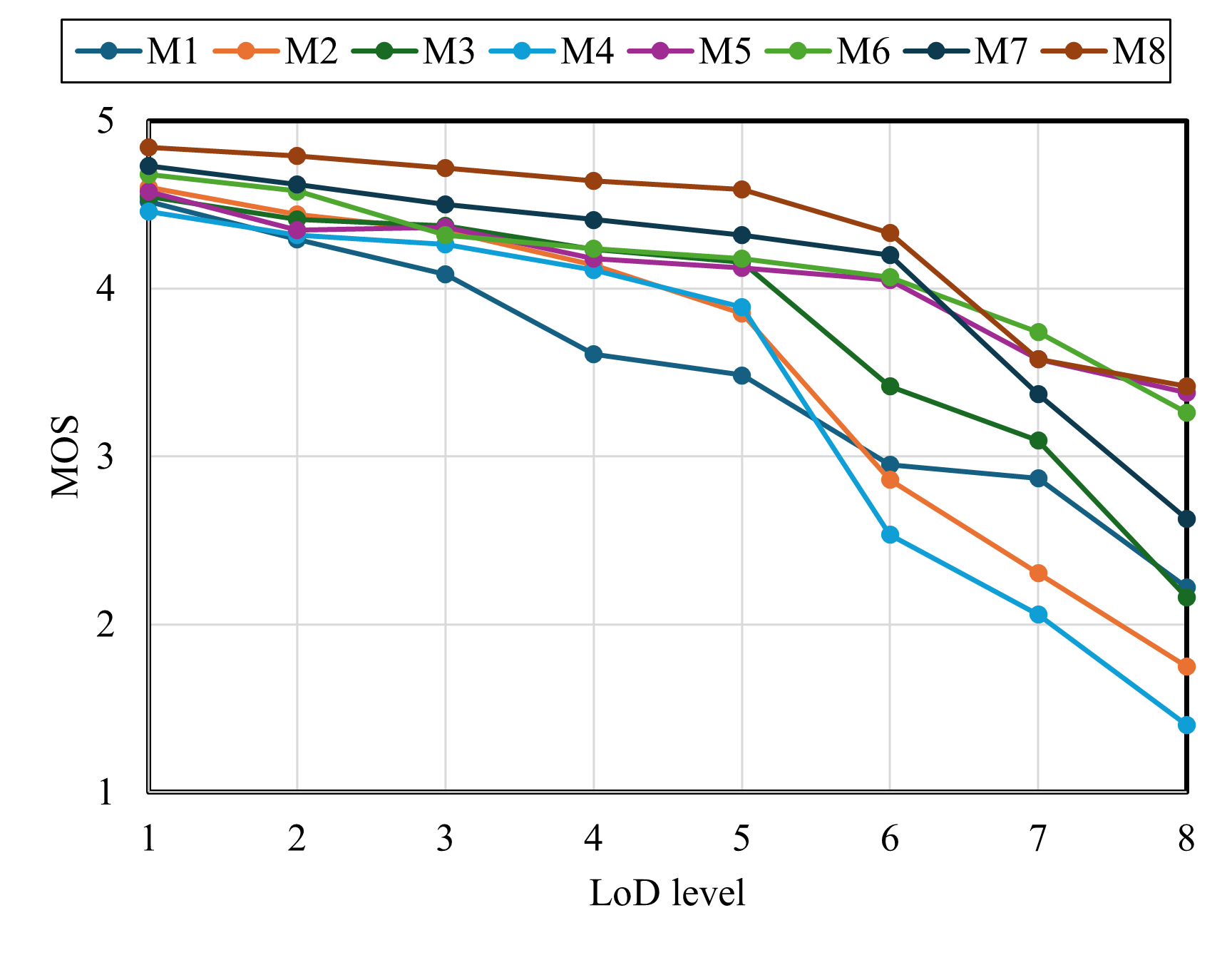}
			}
			\caption{MOSs at different levels of detail across five viewing distances (a)$\sim$(e), and average across all viewing distances (f)}
			\label{fig:LOD_impact}
		\end{figure*}
        \begin{figure}[t]
            \centering
            \includegraphics[width=\columnwidth]{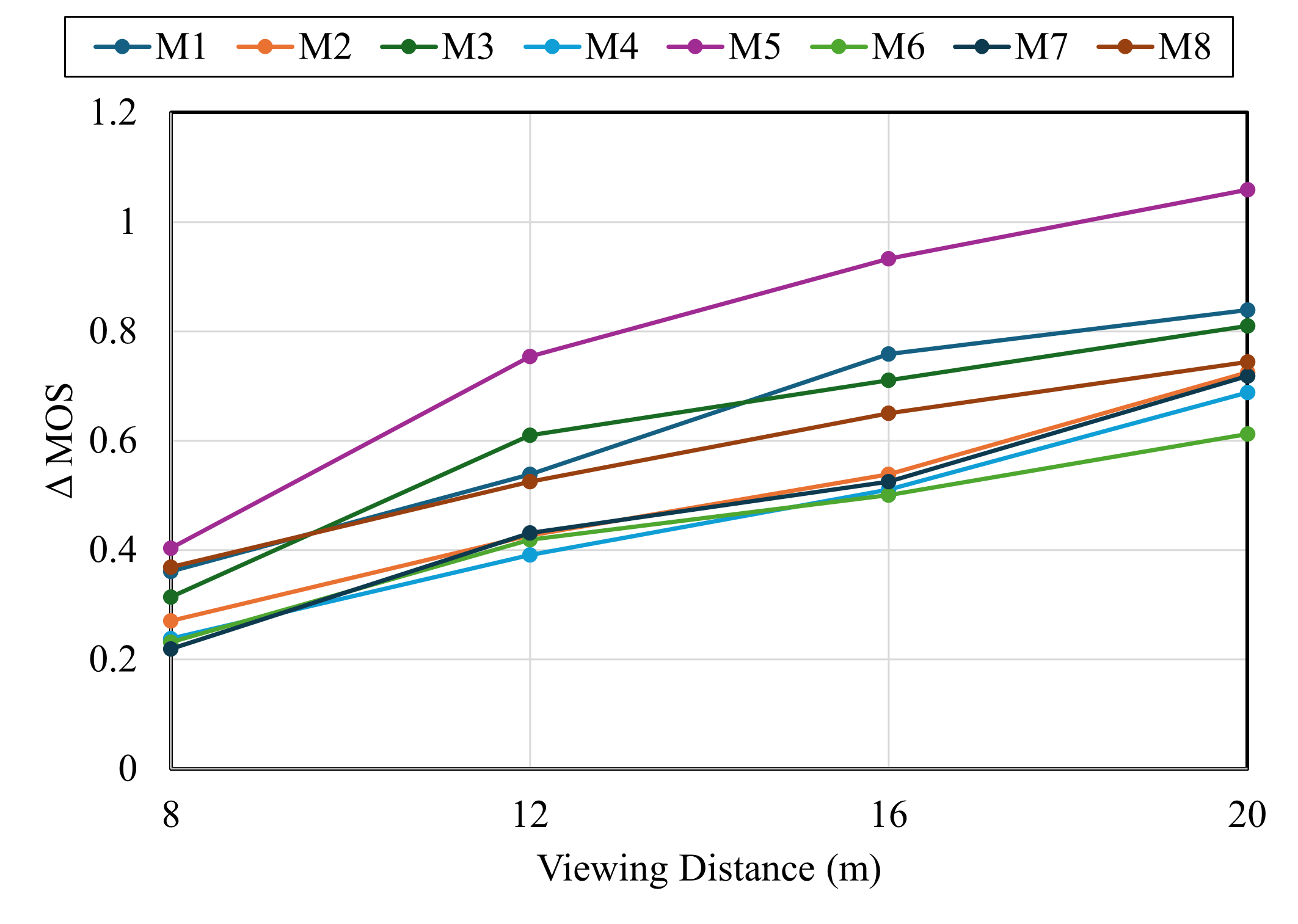}
            \caption{MOS increase amount compared to the value at d=4m of all meshes.}
            \label{fig:Distance_impact}
        \end{figure}


\section{Result Analysis}\label{sec:result_analysis}
\subsection{Statistical Analysis}
Figure~\ref{fig:ci} shows the 95\% confidence interval (CI) and MOS values of the test stimuli in our evaluation. It can be seen that the MOS values cover the entire range from 1 to 5. Also, the CI values are smaller than 0.4. A one-way analysis of variance (ANOVA)~\cite{ANOVA} was conducted to examine whether there were statistically significant differences in MOSs across the considered 3D meshes. The results indicated a significant overall effect, with a p-value $< 0.05$, clearly rejecting the null hypothesis that all mesh means are equal. To identify which specific group pairs exhibited significant differences, Tukey’s Honestly Significant Difference (HSD) post-hoc test~\cite{Tukey} was performed, and the results are reported in Table~\ref{tab:pair-wise-p-value}. The post-hoc analysis revealed that mesh M1 had a significantly different mean score compared to most meshes ($p < 0.05$), but did not differ considerably from the M2 ($p = 1.0$) or M4 meshes ($p = 0.137$). Additionally, no significant differences were observed between the means of the M2 and M4 meshes ($p = 0.130$), M5 and M7 meshes ($p = 0.947$), or M6 and M7 meshes ($p = 0.522$). 
In contrast, the M3 and M8 meshes showed significant differences in mean values when compared with all other meshes ($p < 0.05$).
\begin{table}[]
\caption{Pair-wise p-value from the Tukey HSD test.}
\label{tab:pair-wise-p-value}
\begin{tabular}{|l|l|l|l|l|l|l|l|}
\hline
 & M1 & M2 & M3 & M4 & M5 & M6 & M7 \\ \hline
M2 & 1.000 &  &  &  &  &  &  \\ \hline
M3 & 0.000 & 0.000 &  &  &  &  &  \\ \hline
M4 & 0.137 & 0.130 & 0.000 &  &  &  &  \\ \hline
M5 & 0.000 & 0.000 & 0.000 & 0.000 &  &  &  \\ \hline
M6 & 0.000 & 0.000 & 0.000 & 0.000 & 0.044 &  &  \\ \hline
M7 & 0.000 & 0.000 & 0.000 & 0.000 & 0.947 & 0.522 &  \\ \hline
M8 & 0.000 & 0.000 & 0.000 & 0.000 & 0.000 & 0.034 & 0.000 \\ \hline
\end{tabular}
\end{table}
\subsection{Impacts of Level of Detail and Viewing Distance on MOS}
In this part, we analyze the impact of the level of detail and viewing distance on user perception of dynamic 3D meshes. Fig.~\ref{fig:LOD_impact} shows the MOSs of LoDs of the considered 3D meshes across five viewing distances. It can be seen that the MOS score of all 3D meshes decreases as the level of detail decreases. When 20\% of the mesh's faces are removed (i.e., LoD1), the MOSs of all meshes are higher than 4.5 at distances $d\geq8m$, and higher than 4.0 at $d=4m$. When 40\% of the mesh faces are removed (i.e., LoD2), the MOSs are slightly reduced but still higher than 4.0 at distances $d\geq 8m$ and higher than 3.5 at $d=4m$. Even when half of the original mesh's faces are removed (LoD3), the MOS score is still higher than 4.0 for $d\geq12m$, and higher than 3.5 for $d=4m$ and $d=8m$. These results imply that it is possible to reduce half of a mesh's faces without causing significant degradation in user QoE.

It can also be seen that the effect of reducing the level of detail varies significantly across 3D meshes. At $d=4m$, the decrease level in MOS of the M1 mesh is much higher than that of other meshes from LoD1 to LoD5. As shown in Fig.~\ref{fig:mesh_SI} and Table~\ref{tab:LoD_version_vertice}, the M1 mesh has the highest color complexity and the lowest number of faces. The MOSs of the M4 mesh, which has a similar number of faces but a lower color complexity, are much higher than those of M1 for LoD1$\sim$LoD5. Interestingly, at a lower level of detail from LOD6 to LOD8, the MOS of the M4 mesh becomes lower than that of the M1 mesh. These results indicate that the effect of reducing the level of detail is dependent not only on the number of faces, but also on the complexity of the mesh. 

It can also be seen that the meshes M5, M6, M7, and M8 are less affected by LoD reduction than other meshes at a low level of detail. The MOSs of these meshes are still higher than 3.5 when 80\% of the mesh's faces are removed (i.e., LoD6). Meanwhile, at the same level of detail, the MOSs of the M1, M2, M3, and M4 meshes are lower than 2.5 at $d=4m$. The MOSs of the M5, M6, M7, and M8 meshes drop below 3.5 only when 90\% of mesh faces are removed (LoD7).

The amount of increase in MOS score compared to the value at distance $d=4m$ of all 3D meshes is shown in Fig.~\ref{fig:Distance_impact}. It can be seen that the MOS score increases as the viewing distance increases from $d=4m$ to $d=20m$. This result indicates that the viewing distance affects the user's quality perception. The amount of increase, however, varies across the considered meshes. For all meshes except M5, the MOS is increased by $0.6\sim0.8$ as the viewing distance increases from $d=4m$ to $d=20m$. The increased amount with the M5 mesh, which has the lowest complexity (Fig.~\ref{fig:mesh_SI}), is significantly higher than that of the other meshes. In particular, the MOS of M5 at $d=20m$ is 1.05 higher than the value at $d=4m$. This result indicates that the influence of the viewing distance also relates to the complexity of a 3D mesh.

\subsection{Correlation with Objective Quality Metrics}
		In this part, we assess the correlation between subjective MOS and objective quality metrics. In this paper, we consider three types of metrics as follows.
		\begin{itemize}
			\item \textbf{Image-based metrics}: The image-based metrics include Peak signal-to-noise ratio (PSNR)~\cite{PSNR}, Structural Similarity Index (SSIM)~\cite{SSIM}, Multi-scale Structural Similarity (MS-SSIM)~\cite{MS-SSIM}, Natural Image Quality Evaluator (NIQE)~\cite{NIQUE}, Perception-based Image Quality Evaluator (PIQE)~\cite{PIQE}. To calculate these metrics, we first rendered the 3D meshes at different LoDs and viewing distances. The rendered images of each version were compared to that of the original 3D mesh in the case of full-reference metrics (PSNR, SSIM, and MS-SSIM), and directly evaluated without a reference for no-reference metrics (NIQE and PIQE). 
			\item \textbf{3D-based metrics}: The 3D metrics include Hausdorff Distance (HD), Root Mean Square Error (RMSE), and Chamfer Distance (CD), which are computed directly on 3D meshes~\cite{PCL}.
            \item \textbf{Screen-based metrics}: For the screen-based metrics, we consider the screen-space error (SSE) that measures the error between the distorted and original mesh in pixels on the projected screen~\cite{Hoppe1998}.
		\end{itemize}
		\begin{table}[t]
\centering
\caption{Correlation between the objective quality metrics and subjective scores in terms of PLCC and RMSE.}
\label{tab:metric_evaluation}
\begin{tabular}{|c|l|l|l|}
\hline
\multirow{2}{*}{\textbf{Type}} & \multicolumn{1}{c|}{\multirow{2}{*}{\textbf{Objective Metric}}} & \multicolumn{1}{c|}{\multirow{2}{*}{\textbf{PLCC}}} & \multirow{2}{*}{\textbf{RMSE}} \\
 & \multicolumn{1}{c|}{} & \multicolumn{1}{c|}{} &  \\ \hline
\multirow{5}{*}{Image-based} & PSNR & 0.58 & 0.73 \\ \cline{2-4} 
 & SSIM & 0.49 & 0.81 \\ \cline{2-4} 
 & MS-SSIM & 0.48 & 0.8 \\ \cline{2-4} 
 & NIQE & 0.26 & 0.86 \\ \cline{2-4} 
 & PIQE & 0.19 & 0.88 \\ \hline
\multirow{3}{*}{3D-based} & Hausdorff Distance & 0.33 & 0.84 \\ \cline{2-4} 
 & RMSE & 0.33 & 0.86 \\ \cline{2-4} 
 & Chamfer Distance & 0.49 & 0.78 \\ \hline
Screen-based & SSE & 0.58 & 0.87 \\ \hline
\end{tabular}
\end{table}
		Table~\ref{tab:metric_evaluation} shows the Pearson correlation coefficient (PLCC) and root mean squared error (RMSE) between MOS and the considered objective quality metrics. It can be seen that all considered objective metrics have very low PLCC ($<0.6$) and very high RMSE ( $>0.7$). This indicates that conventional objective metrics have low correlation with subjective scores. This can be explained by the fact that image-based and 3D-based metrics do not consider the viewing distance between the user and the 3D mesh. Meanwhile, our previous analysis shows that the viewing distance strongly affects the user quality perception. Although the screen-space error (SSE) metric includes the viewing distance, the impacts of content characteristics, such as the number of faces, are not considered.

\section{QoE Prediction Model}\label{sec:QoE_model}




As pointed out in previous sections, the impacts of the level of detail and viewing distance vary significantly across different types of 3D meshes. Also, common objective quality metrics have low correlation with subjective scores. In this study, we propose a novel QoE model for predicting the Quality of Experience (QoE) of a 3D mesh at a given viewing distance and level of detail. The model is a random forest tree regression model that combines multiple decision trees to achieve robust and accurate predictions. Specifically, our model takes five input features: the number of faces ($f$), viewing distance ($d$), LoD~($\ell$), geometry spatial information ($s^{geo}$), and color spatial information ($s^{col}$) and outputs the predicted Mean Opinion Score (MOS) as,
\begin{equation}
\textit{QoE} = f_{RF}(\mathbf{x})
\end{equation}, 
where $\mathbf{x} = [f, d, \ell,s^{geo},s^{col}]^T$ represents the feature vector. The spatial information (SI) measures the spatial complexity of the rendered image of a 3D mesh. To obtain the SI value, we apply the method proposed in~\cite{nehme2023textured} in which the rendered image is first converted to grayscale, and then the Sobel filter is applied to compute the edge map.


Bootstrap aggregating (bagging) is used to construct an ensemble of 100 independent decision trees, denoted as $\{T_1, T_2, \ldots, T_{100}\}$. Each tree $T_i$ is trained on a bootstrap sample $\mathcal{D}_i$ obtained by randomly sampling $K$ instances with replacement from the original training dataset of size $K$, where approximately 63.2\% of unique samples appear in each bootstrap sample while the remaining 36.8\% form the out-of-bag (OOB) samples. This bootstrap sampling strategy introduces diversity among the trees while maintaining the statistical properties of the original dataset. The final prediction is computed as the arithmetic mean of predictions from all individual trees:
\begin{equation}
    f_{RF}(\mathbf{x}) = \frac{1}{100}\sum_{i=1}^{100} T_i(\mathbf{x})
\end{equation}
, where $T_i(\mathbf{x})$ represents the prediction of the $i$-th tree. This averaging mechanism reduces prediction variance and improves stability compared to single decision tree models.

During tree construction, each node split is determined by selecting the optimal feature and threshold that minimizes the mean squared error (MSE) of the resulting child nodes. To further enhance diversity and prevent correlation among trees, at each node split, only a random subset of $m_{try}$ features is considered as split candidates, where $m_{try} = \lfloor\sqrt{p}\rfloor = 1$ for our three input features. This random feature selection ensures that even if certain features are highly predictive, different trees will make different splitting decisions, leading to an ensemble of decorrelated predictors. Each tree is grown to its maximum depth without pruning, allowing individual trees to capture complex non-linear relationships in the data, while the ensemble aggregation prevents overfitting.

The training data, comprising 320 samples representing different combinations of mesh configurations, viewing distances, and LoDs, is randomly split into 80\% for training (256 samples) and 20\% for testing (64 samples). The data split is repeated 10 times, and the average result is computed. The model performance is evaluated using multiple complementary metrics: Root Mean Squared Error (RMSE), Pearson Correlation Coefficient (PLCC), Spearman Rank Order Correlation Coefficient (SROCC), and Kendall Rank Order Correlation Coefficient (KROCC). In addition, for each feature $j$, the importance is computed by aggregating the reduction in MSE achieved when splitting on that feature across all nodes in all trees, weighted by the proportion of samples reaching each node. The importance of feature $j$ is defined as: 
\begin{equation}
    \text{Imp}(j) = \frac{1}{100}\sum_{i=1}^{100}\sum_{k \in K_i(j)} \frac{|S_k|}{K}\Delta\text{MSE}(k, j)
\end{equation}
, where $K_i(j)$ denotes the set of nodes in tree $i$ that split on feature $j$, $|S_k|$ is the number of samples at node $k$, and $\Delta\text{MSE}(k, j)$ represents the MSE reduction at that split.

Table~\ref{tab:model_comparison} presents a performance comparison between the proposed QoE model and two baseline models. The linear regression model uses the same inputs as our proposed model and predicts the MOS as a weighted sum of inputs. The Nguyen's model proposed in~\cite{DucICIP2024} predicts the MOS from two input values of the number of vertices and the viewing distance. The results demonstrate that the proposed QoE model achieves superior performance across all evaluation metrics, with the lowest prediction error (RMSE = 0.19) and highest correlation coefficients (PLCC = 0.98, SROCC = 0.97, KROCC = 0.88). Compared to the linear regression model, the proposed model reduces RMSE by 60\%, indicating its ability to capture non-linear relationships between mesh characteristics and perceived quality. Although considering the viewing distance, Nguyen's model exhibits poor performance on our data, with a PLCC of $0.41$ and a RMSE of $0.80$. This is because the level of detail, which is a key influencing factor on user QoE, is not taken into account.

\begin{table}[t]
\centering
\caption{Performance comparison of the proposed QoE model and reference models for QoE prediction on the test set (averaged over 10 runs).}
\label{tab:model_comparison}
\begin{tabular}{lcccc}
\hline
\textbf{Model} & \textbf{RMSE} & \textbf{PLCC} & \textbf{SROCC} & \textbf{KROCC} \\
\hline
Linear Regression & 0.48 & 0.85 & 0.93 & 0.78 \\
Nguyen's~\cite{DucICIP2024} & 0.80 & 0.41 & 0.48 & 0.34 \\
\textbf{Proposed Model} & \textbf{0.19} & \textbf{0.98} & \textbf{0.97} & \textbf{0.88} \\
\hline
\end{tabular}
\end{table}

\begin{table}[t]
\centering
\caption{Feature importance scores}
\label{tab:feature_importance}
\begin{tabular}{lc}
\hline
\textbf{Feature} & \textbf{Importance} \\
\hline
LoD ($\ell$) & 0.636 \\
Number of Faces ($f$) & 0.107 \\
Viewing Distance ($d$) & 0.124\\
Geometry Spatial Information ($s^{geo}$) & 0.059 \\
Color Spatial Information ($s^{col})$ & 0.074\\
\hline
\end{tabular}
\end{table}

Table~\ref{tab:feature_importance} shows the feature importance scores computed from the proposed model based on mean decrease in impurity. The results reveal that the LoD is the most influential feature, accounting for 63.6\% of the total importance, indicating that the level of detail has the strongest impact on perceived mesh quality. The viewing distance has the second-highest importance score of 12.4\%. Among features representing content characteristics, the number of faces has the highest importance of 10.7\%, followed by the color spatial information (7.4\%), and the geometry spatial information (5.9\%). These importance scores provide valuable insights into the perceptual factors driving QoE and can guide optimization strategies for mesh rendering applications.

\section{QoE-aware Resource Allocation Framework}\label{sec:QoE_aware_resource_allocation}
In this part, we present a QoE-aware resource allocation framework that utilizes the proposed QoE model (Section~\ref{sec:QoE_model}) to optimize user experience under a resource-constrained scenario. Assume that a VR application with multiple dynamic 3D meshes operates under a resource-constrained environment. Multiple levels of detail are available for each mesh. The problem is to decide which LoD should be selected for each 3D mesh to maximize the overall user experience while satisfying the resource constraints.

\subsection{Problem Formulation}

We formulate the LOD selection problem as a binary integer programming (BIP) optimization to maximize the overall QoE while respecting computational budget constraints. Consider a scene with $N$ 3D mesh models, where each model $n$ can be rendered at one of $M$ possible Levels of Detail (LOD). Let $x_{nm} \in \{0,1\}$ denote a binary decision variable indicating whether LoD $m$ is selected for mesh $n$. Each LoD $m$ for mesh $n$ has a corresponding number of faces $f_{nm}$ and viewing distance $d_n$. The QoE of mesh $n$ at LoD $m$ is predicted using the QoE model presented in Section~\ref{sec:QoE_model}: $\textit{QoE}_{nm} = \hat{f}_{RF}(f_{nm}, d_n, \ell_m, s^{geo}_{n}, s^{col}_n)$, where $\ell_m$ represents the LOD percentage, and $s^{geo}_{n}$ and $s^{col}_n$ are respectively the geometry spatial information and color spatial information.

The optimization problem is formulated as:

\begin{align}
\max_{x_{nm}} \quad & \sum_{n=1}^{N} \sum_{m=1}^{M} \textit{QoE}_{nm} \cdot x_{nm} \label{eq:objective} \\
\text{subject to} \quad & \sum_{n=1}^{N} \sum_{m=1}^{M} f_{nm} \cdot x_{nm} \leq B \label{eq:budget} \\
& \sum_{m=1}^{M} x_{nm} = 1, \quad \forall n \in \{1, \ldots, N\} \label{eq:selection} \\
& x_{nm} \in \{0, 1\}, \quad \forall n, m \label{eq:binary}
\end{align}
where equation~\eqref{eq:objective} maximizes the total QoE across all meshes, equation~\eqref{eq:budget} ensures that the total face count does not exceed the computational budget $B$, equation~\eqref{eq:selection} guarantees that exactly one LoD is selected for each mesh, and equation~\eqref{eq:binary} enforces binary decision variables. The computational budget $B$ is represented by the total number of faces of 3D meshes in the scene.
\begin{algorithm}[t]
\caption{LoD Selection Algorithm}
\label{alg:branch_and_bound}
\begin{algorithmic}[1]
\Require Binary integer program with objective $\max \sum_{n,m} \textit{QoE}_{nm} x_{nm}$, constraints~\eqref{eq:budget}--\eqref{eq:binary}
\Ensure Optimal LOD allocation $\mathbf{x}^* = \{x_{nm}^*\}$
\State Initialize priority queue $\mathcal{Q} \leftarrow \{\text{root node with all } x_{nm} \in [0,1]\}$
\State Initialize incumbent solution $\mathbf{x}_{\text{best}} \leftarrow \emptyset$, best value $z_{\text{best}} \leftarrow -\infty$
\While{$\mathcal{Q} \neq \emptyset$}
    \State \textbf{Select:} Remove node $\mathcal{N}$ with highest upper bound from $\mathcal{Q}$
    \State \textbf{Relax:} Solve Linear Programming (LP) relaxation of subproblem at $\mathcal{N}$ to obtain solution $\mathbf{x}_{\text{LP}}$ and value $z_{\text{LP}}$
    \If{LP is infeasible \textbf{or} $z_{\text{LP}} \leq z_{\text{best}}$}
        \State \textbf{Prune:} Discard node $\mathcal{N}$ \Comment{Fathom by infeasibility or bound}
        \State \textbf{continue}
    \EndIf
    \If{$\mathbf{x}_{\text{LP}}$ is integer-feasible} \Comment{All $x_{nm} \in \{0,1\}$}
        \State \textbf{Update incumbent:} $\mathbf{x}_{\text{best}} \leftarrow \mathbf{x}_{\text{LP}}$, $z_{\text{best}} \leftarrow z_{\text{LP}}$
    \Else
        \State \textbf{Branch:} Select fractional variable $x_{ij}$ where $x_{ij} \notin \{0,1\}$
        \State Create child node $\mathcal{N}_0$ with $x_{ij} = 0$ fixed, add to $\mathcal{Q}$
        \State Create child node $\mathcal{N}_1$ with $x_{ij} = 1$ fixed, add to $\mathcal{Q}$
    \EndIf
\EndWhile
\State \Return $\mathbf{x}_{\text{best}}$ as optimal solution with value $z_{\text{best}}$
\end{algorithmic}
\end{algorithm}
\begin{table*}[t]
\centering
\caption{Performance comparison of the proposed QoE-aware allocation method and two reference methods. (averaged over 10 runs)}
\label{tab:optimization_comparison}
\begin{tabular}{ccccccccccc}
\hline
\multirow{2}{*}{\textbf{Budget}} & \multicolumn{3}{c}{\textbf{Proposed}} & \multicolumn{3}{c}{\textbf{Greedy}} & \multicolumn{3}{c}{\textbf{Equal Dist.}} \\
\cmidrule(lr){2-4} \cmidrule(lr){5-7} \cmidrule(lr){8-10}
& \textbf{QoE} & \textbf{Usage \%} & \textbf{Time (ms)} & \textbf{QoE} & \textbf{Usage \%} & \textbf{Time ($\mu$s)} & \textbf{QoE} & \textbf{Usage \%} & \textbf{Time ($\mu$s)} \\
\hline
25,000  & 21.04 & 99.32 & 1.60 & 20.33 & 98.71 & 12.92 & 16.12 & 43.95 & 9.37 \\
50,000  & 27.85 & 98.86 & 3.91 & 24.03 & 98.67 & 12.56 & 23.31 & 58.48 & 7.58 \\
75,000  & 30.31 & 99.15 & 6.77 & 28.08 & 99.43 & 11.56 & 26.03 & 60.78 & 9.06 \\
100,000 & 32.14 & 99.08 & 2.41 & 30.65 & 99.04 & 11.18 & 29.24 & 63.63 & 6.96 \\
125,000 & 33.22 & 98.86 & 3.62 & 31.27 & 99.27 & 10.44 & 31.15 & 71.79 & 6.32 \\
150,000 & 34.07 & 99.24 & 3.63 & 32.35 & 98.66 & 9.87 & 31.27 & 62.01 & 6.63 \\
175,000 & 34.93 & 99.08 & 2.04 & 33.69 & 99.83 & 10.18 & 32.52 & 67.77 & 6.87 \\
200,000 & 35.43 & 98.95 & 3.17 & 33.98 & 97.37 & 9.61 & 33.26 & 63.20 & 6.53 \\
225,000 & 35.79 & 99.72 & 1.99 & 34.49 & 91.26 & 8.80 & 33.47 & 62.06 & 6.53 \\
250,000 & 36.00 & 99.08 & 1.16 & 36.00 & 99.08 & 9.32 & 33.69 & 58.97 & 6.10 \\
275,000 & 36.13 & 99.30 & 1.71 & 36.05 & 97.78 & 8.75 & 33.83 & 56.04 & 6.06 \\
300,000 & 36.28 & 98.03 & 1.19 & 36.26 & 96.69 & 8.20 & 33.83 & 51.37 & 6.15 \\
\hline
\end{tabular}
\end{table*}
\subsection{LoD Selection Method}

The proposed solution obtains the globally optimal LOD allocation by solving the binary integer programming problem using the branch-and-bound algorithm~\cite{BrandAndBoundAlg}. This method guarantees global optimality by systematically exploring the solution space while pruning suboptimal branches. The algorithm utilizes linear programming relaxations at each node of the branch-and-bound tree and employs cutting plane methods to refine the bounds. This approach identifies the allocation that maximizes total QoE while strictly adhering to the budget constraint. The pseudo-code of the algorithm is summarized in Algorithm~\ref{alg:branch_and_bound}.

\subsection{Experimental Setup}

The evaluation is conducted using a dataset of $N = 8$ meshes with varying complexities, ranging from 4,063 to 211,874 faces. There are $M = 8$ LoDs for each meshes as described in Table~\ref{tab:LoD_version_vertice}. Viewing distance of each mesh is randomly sampled from $\{4, 8, 12, 16, 20\}$ meters. The evaluation is carried out across 12 different budget levels ranging from 25,000 to 300,000 faces, corresponding to approximately 5\% to 60\% of the total maximum face count (if all meshes were rendered at 80\% LOD). For each budget level and allocation method, we compute: (1) the total predicted QoE summed across all meshes, (2) the budget utilization percentage, and (3) the decision time in seconds. The viewing distance selection is repeated 10 times, and the average results are reported.

We evaluate the proposed method against the two reference methods: \textbf{Greedy} and \textbf{Equal Distribution}. The greedy method starts by allocating the minimum LOD (LoD8) to all meshes to ensure feasibility. It then greedily upgrades each mesh sequentially to the highest possible LoD that fits within the remaining budget. The equal distribution method divides the budget equally among all meshes, allocating $B/N$ faces to each mesh. For each mesh $n$, the highest LoD that fits within the allocated budget is selected.
\subsection{Results and Comparison}

Table~\ref{tab:optimization_comparison} presents a comparison of the proposed method against the two reference methods across different budget levels. It can be seen that the proposed method consistently achieves the highest total QoE across all budget levels, demonstrating improvements of 0.2--15.9\% over the greedy method and 6.5--30.5\% over the equal distribution method at constrained budgets. At the lowest budget (25,000 faces), where resource constraints are most severe, the proposed method achieves 3.5\% higher QoE than the greedy approach and 30.5\% higher than equal distribution, demonstrating its ability to optimize resource allocation even in highly constrained scenarios.

Both the proposed method and the greedy method exhibit high budget utilization rates (91.3--99.8\%), effectively using available resources without exceeding constraints. In contrast, the equal distribution method exhibits very low utilization at constrained budgets (43.95\% at 25,000 faces) due to excluding large meshes that cannot fit the minimum LOD within their equal budget share. At medium budgets (50,000--125,000 faces), utilization improves to 58--72\%, but remains substantially below optimal.

The proposed method requires less than 7.0 milliseconds per budget level, demonstrating reasonable computational overhead despite solving a complex integer programming problem with 64 binary variables ($8$ meshes $\times$ $8$ LoDs). Both reference methods exhibit similar performance: the greedy method completes in 8--13 microseconds, while the equal distribution method requires 6--10 microseconds.

\section{Conclusions}\label{sec:conclusion}
In this paper, we conduct a large-scale subjective evaluation of dynamic 3D meshes under different levels of detail and viewing distances in Virtual Reality. Experimental results reveal that users' quality perception is affected not only by the level of detail and viewing distance, but also by content characteristics, including the number of faces and complexity level. Our analysis also found that common objective quality metrics have very low correlation with subjective scores. To address this problem, we develop a novel QoE model that can accurately predict the MOS score of a dynamic 3D mesh given a level of detail and viewing distance. The proposed QoE model is leveraged to develop a QoE-aware resource allocation framework for 3D meshes in a VR environment. The proposed framework can significantly improve the overall user QoE under resource-constrained scenarios. In future work, we will develop methods for automatic generation of the Level of Detail of dynamic 3D meshes. Also, the findings in this study will be applied to optimize the transmission and selection of 3D meshes in VR applications.

\bibliographystyle{IEEEbib}
\bibliography{myref}


\end{document}